\documentclass[conference]{IEEEtran}
\IEEEoverridecommandlockouts
\usepackage{amsmath,amssymb,amsfonts}
\usepackage{physics}
\usepackage{algorithmic}
\usepackage{graphicx}
\usepackage{textcomp}
\usepackage{hyperref}
\usepackage{enumitem}
\usepackage{fancyhdr}
\usepackage{ragged2e}
\usepackage{floatrow}
\usepackage{xcolor}
\usepackage{listings}
\usepackage{qcircuit}
\def\BibTeX{{\rm B\kern-.05em{\sc i\kern-.025em b}\kern-.08em
    T\kern-.1667em\lower.7ex\hbox{E}\kern-.125emX}}
\usepackage[
backend=biber,
style=numeric,
url=false,
isbn=false,
eprint=false,
doi=false,
sorting=none
]{biblatex}
\newcommand{\doiorurl}{%
  \iffieldundef{doi}
    {\iffieldundef{url}
       {}
       {\strfield{url}}}
    {http://dx.doi.org/\strfield{doi}}%
}

\newcommand{\myhref}[1]{%
 \ifboolexpr{%
   test {\ifhyperref}
   and
   not test {\iftoggle{bbx:url}}
   and
   not test {\iftoggle{bbx:doi}}
  }
  {\href{\doiorurl}{#1}}
  {#1}
}

\DeclareFieldFormat{title}{\myhref{\mkbibemph{#1}}}
\DeclareFieldFormat
  [article,inbook,incollection,inproceedings,patent,thesis,unpublished]
  {title}{\myhref{\mkbibquote{#1\isdot}}}
\newcommand{\register}{\texttt}
\newcommand{\tiasm}{\textsc}

\author{

    \IEEEauthorblockN{
        Tobias Schmale\IEEEauthorrefmark{1}\textsuperscript{1}, 
        Bence Temesi\IEEEauthorrefmark{1}\textsuperscript{2}, 
        Alakesh Baishya\IEEEauthorrefmark{1},
        Nicolas Pulido-Mateo\IEEEauthorrefmark{2}\IEEEauthorrefmark{3},
        Ludwig Krinner\IEEEauthorrefmark{2}\IEEEauthorrefmark{3},\\
        Timko Dubielzig\IEEEauthorrefmark{2},
        Christian Ospelkaus\IEEEauthorrefmark{2}\IEEEauthorrefmark{3},
        Hendrik Weimer\IEEEauthorrefmark{1},
        Daniel Borcherding\IEEEauthorrefmark{1}\textsuperscript{3}
    }\\
    \IEEEauthorblockA{
        \IEEEauthorrefmark{1}\textit{Institut für Theoretische Physik, }
        \textit{Leibniz Universität Hannover, }
        Appelstraße 2, 30167 Hannover, Germany 
    }
    \IEEEauthorblockA{
        \IEEEauthorrefmark{2}\textit{Institut für Quantenoptik, }
        \textit{Leibniz Universität Hannover, }
        Welfengarten 1, 30167 Hannover, Germany
    }
    \IEEEauthorblockA{
        \IEEEauthorrefmark{3}\textit{Physikalisch-Technische Bundesanstalt, }
        Bundesallee 100, 38116 Braunschweig, Germany
    }
}

\begin{document}
\fancypagestyle{copyright}{\fancyhf{}\renewcommand{\headrulewidth}{0pt}\fancyfoot[C]{\vspace{-0.7cm} \justifying \small \noindent \copyright 2022 IEEE. PERSONAL USE OF THIS MATERIAL IS PERMITTED. PERMISSION FROM IEEE MUST BE OBTAINED FOR ALL OTHER USES, IN ANY
CURRENT OR FUTURE MEDIA, INCLUDING REPRINTING/REPUBLISHING THIS MATERIAL FOR ADVERTISING OR PROMOTIONAL PURPOSES, CREATING
NEW COLLECTIVE WORKS, FOR RESALE OR REDISTRIBUTION TO SERVERS OR LISTS, OR REUSE OF ANY COPYRIGHTED COMPONENT OF THIS
WORK IN OTHER WORKS}}

\title{Backend compiler phases for trapped-ion \\quantum computers}


\maketitle


\begin{abstract}
A promising architecture for scaling up quantum computers based on trapped ions are so called Quantum Charged-Coupled Devices (QCCD). These consist of multiple ion traps, each designed for solving specific tasks, that are connected by transport links.
In this paper we present the backend compiler phases needed for running quantum circuits on a QCCD architecture, while providing strategies to solve the optimization problems that occur when generating assembly instructions. We implement and test these strategies for the QVLS-Q1 chip architecture. 

\end{abstract}


\footnotetext[1]{tobias.schmale@itp.uni-hannover.de}
\footnotetext[2]{bence.temesi@itp.uni-hannover.de}
\footnotetext[3]{daniel.borcherding@itp.uni-hannover.de}

\thispagestyle{copyright}
\section{Introduction}
Trapped ions are a promising candidate for realizing universal quantum processors.  This is due to their high-fidelity state preparation, readout and universal gate operations as well as their long qubit coherence times \cite{bruzewicz2019}. 


Several architectures have been proposed for scaling up quantum devices based on trapped ions. In the simplest architecture, ions are arranged in a 1D string with a shared potential, where gates are implemented by addressing individual ion states and shared motional modes using lasers and microwaves \cite{bruzewicz2019}. However, it was shown that performing high-fidelity two-qubit gates becomes impractical for more than approximately 100 ions \cite{bruzewicz2019}. To overcome these challenges, an architecture, called Quantum Charge-Coupled Design (QCCD) was proposed, which consists of multiple ion traps optimized for their specific function as well as transportation links between them \cite{wineland1998experimental, kielpinski2002}. The first implementation of a QCCD architecture can be found in \cite{pino2020demonstration}. See \cite{bruzewicz2019} for a review of trapped ion architectures. 

Similar to classical computers, a compiler is needed to run algorithms on hardware. In this case, the compiler needs to turn quantum circuits into concrete sequences of ion shuttle operations and gate executions. The main challenge hereby is to simultaneously optimize the order of quantum gates and the number of ion shuttles while maintaining an appropriate compilation time. 

\begin{figure}
    \centering
    \includegraphics[width=0.7\textwidth]{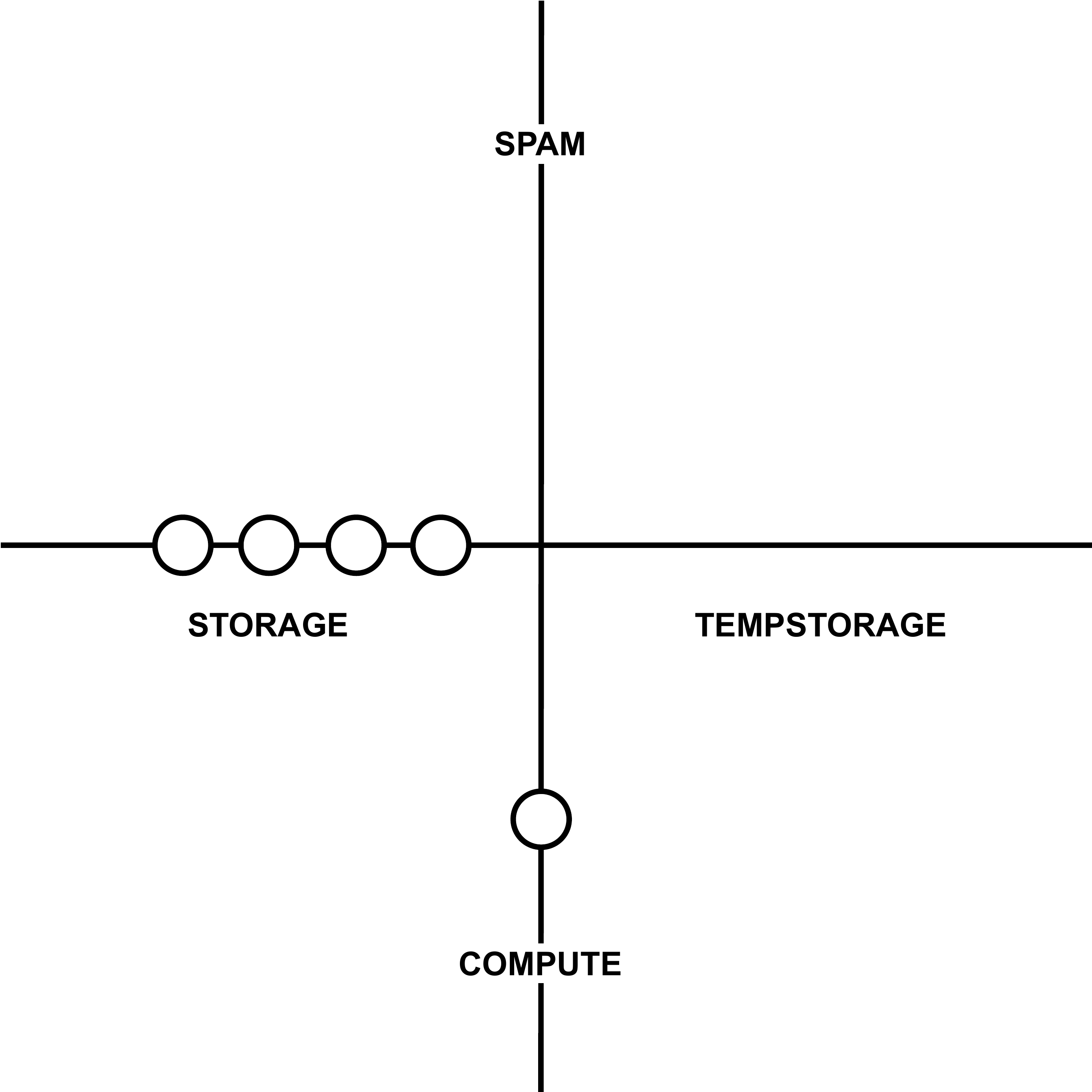}
    \caption{Simplistic illustration of the QVLS-Q1 chip architecture from a top view showing the \register{STORAGE}, \register{TEMPSTORAGE}, \register{SPAM} and \register{COMPUTE} register connected by an X-junction. In the depicted configuration four ions are stored in the \register{STORAGE} register and one is located in the \register{COMPUTE} register. Movements of ions are only allowed along the transportation links (black lines).}
    \label{fig:qvls_q1_chip}
\end{figure}

In the following, we present backend compilation phases for a trapped-ion QCCD architecture and propose strategies to optimize gate executions and ion movements. While these strategies are applicable to general QCCD architectures we demonstrate our approach using a new QCCD architecture (see Fig.~\ref{fig:qvls_q1_chip}) that is currently being developed by the Quantum Valley Lower Saxony (QVLS) for the QVLS-Q1 quantum computer.

The code developed for this project is available under \cite{tiasm2022compiler}.


\section{Concept}
\subsection{Hardware}



The QCCD architecture developed for the QVLS-Q1 quantum computer carries at most 50 ions and consists of the following registers connected by an X-junction:
\begin{itemize}
    \item \register{SPAM}: State preparation and measurement register with a maximal trap capacity of one ion.
    \item \register{STORAGE} / \register{TEMPSTORAGE}: Storage registers with a maximal trap capacity of 50 ions each.
    \item \register{COMPUTE}: Computational register based on \cite{zarantonello2019robust} with a maximal trap capacity of 2 ions and the possibility to apply the single-qubit gate $$\tiasm{r}(\phi, \theta) = \cos{\phi}\,\text{e}^{\text{i}\theta\sigma_x / 2} + \sin{\phi}\,\text{e}^{\text{i}\theta\sigma_y / 2}\;,$$ where we additionally define $\tiasm{rx}(\theta) \equiv \tiasm{r}(0, \theta)$ and $\tiasm{ry}(\theta)\equiv \tiasm{r}(\pi/2, \theta)$. We can also apply the two-qubit gate $\tiasm{rxx}(\theta) \equiv \text{e}^{\text{i}\theta \sigma_x \otimes \sigma_x / 2}$.\\ Here, $$
\sigma_x =
\begin{pmatrix}0&1\\1&0\end{pmatrix} \quad\text{and}\quad
\sigma_y =
\begin{pmatrix}0&-i\\i&0\end{pmatrix} 
$$ are the Pauli matrices and $\theta$ is some real angle.
\end{itemize}
In this architecture, only the ions closest to the junction can be moved from one register to another following the transportation links. However, directly exchanging neighboring ions within one register is not allowed. See Fig.~\ref{fig:qvls_q1_chip} for a simplistic illustration of the registers and transportation links of the QVLS-Q1 chip. 

Notice also, that in contrast to other QCCD architectures the gates can only be applied to at most two qubits simultaneously in the \register{COMPUTE} register, while all-to-all qubit connectivity is achieved by physically moving ions between the registers. 
The behavior of the registers can be best described in terms of multiple stack data structures: each register on the chip can be understood as a stack where the top element is the ion closest to the junction. Moving an ion from one register to another can then refer to popping and pushing an ion from the corresponding stacks.

\subsection{Assembly Language}

To describe operations on a QCCD architecture, assembly instructions for moving the ions and applying operations on them are needed. For the QVLS-Q1 chip this is realized by our trapped-ion assembly language (TIASM). 
The main components of the language needed for compiling a quantum circuit to TIASM are given by:

\begin{itemize}
    \item $\tiasm{quantum\_register}(n)$: Loads $n$ ions to the \register{STORAGE} register.
    \item $\tiasm{classical\_register}(n)$: Creates a classical register of $n$ bits for storing the measurement results.
	\item $\tiasm{move}(r_1,r_2)$: Moves the ion closest to the junction of register $r_1$ to register $r_2$.
	\item $\tiasm{measure} \rightarrow c$: Applies a measurement to the ion in the \register{SPAM} register and stores the measurement result (0 or 1) in a classical bit with label $c$.
	\item $\tiasm{prepare}(r)$: If register $r = \register{SPAM}$ a specific laser pulse is applied to the ion in the \register{SPAM} register to prepare it in a suitable ion state. If $r = \register{COMPUTE}$ a microwave pulse is applied in the \register{COMPUTE} register to prepare the ion in the initial qubit state referred to as $\ket{0}$.
	\item $\tiasm{rx}(\theta)\ i_{\text{stack}}$ / $\tiasm{ry}(\theta)\ i_{\text{stack}}$ / $\tiasm{r}(\phi, \theta) \ i_{\text{stack}}$: Applies an $\tiasm{rx}(\theta)$, $\tiasm{ry}(\theta)$ or $\tiasm{r}(\phi, \theta)$ gate in the \register{COMPUTE} register to the ion at location $i_{\text{stack}}$ within the stack.
	\item $\tiasm{rxx}(\theta)$: Applies the $\tiasm{rxx}(\theta)$ two-qubit gate to both ions in the \register{COMPUTE} register.
\end{itemize}

See \cite{tiasm2022compiler} for the complete TIASM grammar.


\subsection{Backend compilation phases}

\begin{figure}
    \centering
    \includegraphics[width=0.85\textwidth]{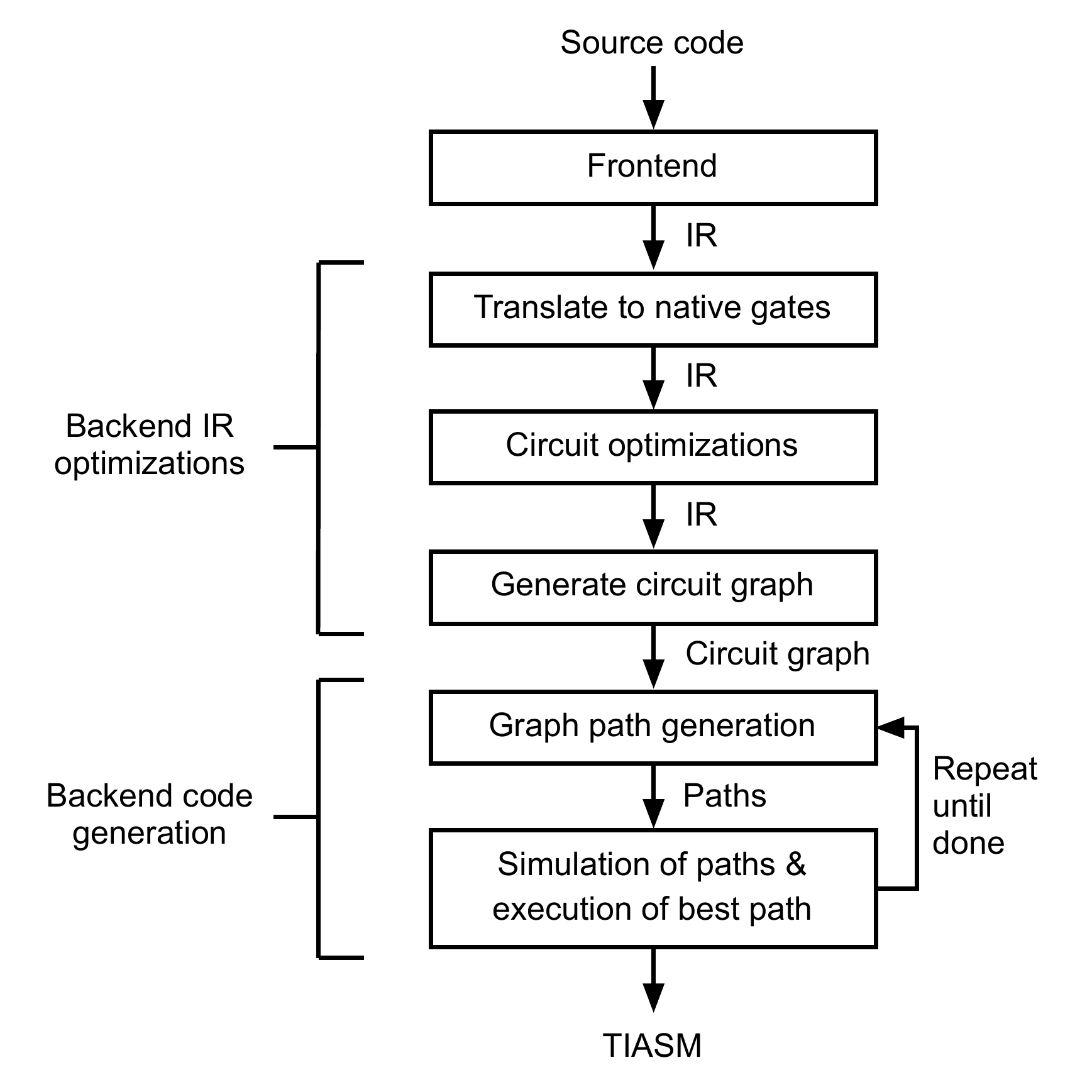}
    \caption{Illustration showing the phases of the compiler. 1. A frontend transforms the source code to the intermediate representation (IR). 2. Optimizations dependent on the underlying hardware are applied to the IR. This includes the translation to native gates as well as circuit optimizations based on gate identities. 3. The assembly instructions are generated by orchestrating the gates and the ions using a circuit graph.}
    \label{fig:compiler_phases}
\end{figure}

The task of the compiler is to translate a high-level quantum programming language \cite{heim2020quantum} to our assembly language TIASM. As for classical compilers this task can be split into a frontend part, which consists of a lexer, parser and intermediate code generator and a backend part, which optimizes the intermediate code and generates the assembly instructions. See Fig.~\ref{fig:compiler_phases} for an overview of the different compilation phases.

In this work we assume that a frontend generating an intermediate representation (IR) exists and focus on the different phases of the backend part of the compiler. In the following we consider IRs based on OpenQASM \cite{cross2017open}. The concepts for the backend compilation part are however also applicable when using different IRs.
The different phases are now described in detail:

\subsubsection{Translate to native gates}
The native gates that can be applied on the QVLS-Q1 chip and on most trapped-ion chips are $\tiasm{rx}(\theta)$, $\tiasm{ry}(\theta)$, $\tiasm{rxx}(\theta)$ \cite{Colin2019}. To convert a quantum circuit consisting of arbitrary quantum gates to the trapped-ion basis gates we are using the existing transpilation methods provided by the Python module called \texttt{pytket} \cite{pytket}.

\subsubsection{Circuit optimizations}
After efficiently unrolling an input circuit to our native set of gates, we continue to use the \texttt{pytket} module for gate-level circuit optimizations. Recent benchmarks comparing different transpilation methods \cite{benchmark} showed that \texttt{pytket} gives on average the fastest transpilation times  due to its C++ core, while resulting in the lowest number of gates at the same time. \texttt{pytket} repeatedly merges single qubit rotations and uses commutative cancellation until the number of gates are optimized.
On top of that, we extend its capabilities with methods unique to our ion-trap machine based on \cite{maslov2017basic}. There, Maslov gives a detailed blueprint on how to reduce circuit depth if one works in the set of $\{\tiasm{rx}, \tiasm{ry}, \tiasm{rxx}\}$ gates. One of the ideas is to use \textit{templates}, which are sequences of gates that evaluate to identity \cite{maslov2008quantum}. A useful circuit identity in our case is given by:
\label{circ:maslov-templ}
\\[\baselineskip]
\resizebox{0.48\textwidth}{!}{
$$ \Qcircuit @C=1em @R=.7em {
      & \gate{\tiasm{rx}(\theta_1)} & \gate{\tiasm{ry}(\theta_2)} & \gate{\tiasm{rx}(\theta_1)} & \gate{\tiasm{r}(\theta_3, \theta_4 - \pi)} & \qw
      } \hspace{1em} =
      \Qcircuit @C=1em @R=.7em {
      & & \qw
}
$$
}
\\[\baselineskip]
where for any $\theta_1, \theta_2 \in \mathbb{R}$ we can compute the corresponding angles of $\theta_3, \theta_4 \in \mathbb{R}$, more details in \cite{maslov2017basic}. See a short example in Appendix \ref{appendix:optimizationexample}. 
Note however, that this method does not yet help with the reduction of two-qubit gates, which are generally more prone to errors than single qubit gates. While implementing the template above has no disadvantages, some of the rest of the strategies from \cite{maslov2017basic} can introduce tradeoffs between circuit runtime and fidelity errors, which is why we omit these strategies here. 

\subsubsection{Generate circuit graph}
After these circuit optimizations, we can proceed with the actual code generation.
Due to the sequential nature in which gates need to be executed on this hardware, it makes sense to group gates based on which qubits they act on. To preserve the non-trivial dependencies created by multi-qubit gates, we generate a directed, acyclic ''gate dependency'' graph (DAG) from the circuit \cite{sakiMuzzle2021}. A valid hardware-realization of the circuit, i.e. \emph{circuit path}, consists of a traversal of this graph, that visits a node only once all parents have been visited.

\subsubsection{Graph serialization}
In general there are many possible circuit paths that satisfy the gate dependencies. By only following the edges in the DAG we generate a subset of those circuit paths that leads to fewer exchanges of the ions in the compute zone.

Following this strategy we generate up to $n_p$ circuit paths of length $n_g$, meaning that each path contains at most $n_g$ \tiasm{rxx} gates. These paths are then simulated using the strategies from the next section and the best one is chosen. The gates from this best path are then deleted from the graph and a next set of paths is generated until all gates have been executed and removed from the graph.

See Appendix \ref{appendix:graphpathgeneration} for an example showing the different graph paths that we generate for a given circuit.


\subsubsection{Ion orchestration} \label{sec:ion_orchestration}
For generating the actual ion movements, we assume that a circuit path, as described in the previous section, is given.
Initially, we assign a qubit id to a physical ion once that qubit appears in a generated circuit path for the first time. That way the initial order in which the ions are placed in the storage register is optimal.
The task then is to move ions on the chip, such that the correct ions end up in the \register{COMPUTE} register in the sequence determined by the path.
To solve this problem, we apply the following heuristics:

\begin{itemize}
    \item Smallest junction distance (\textit{JunctionDistance}): When moving two ions to the compute zone we start with the ion that is closer to the junction.
    \item Compute ordering (\textit{ComputeOrder}): When two ions have the same junction distance we first move the ion to the compute zone that will stay there longer as determined by the next gates in the current circuit path.
    \item Use \register{SPAM} as temporary storage (\textit{SpamStorage}): Whenever there is a free location in the \register{SPAM} area we use
    it to keep upcoming ions in the graph path closer to the junction.
    \item Partner sorting (\textit{PartnerSorting}): From the complete circuit graph we can conclude which ions will need to be paired up in the \register{COMPUTE} area in the future. We refer to these qubits as next "partner qubits". By using the \register{SPAM} and \register{COMPUTE} register we can exploit the information about the next partner qubits to improve the ordering of the ions. Whenever ions are moved from one storage register to another, we try to place these partners closer to each other.
    
\end{itemize}
An example of a graph path being executed using the \textit{JunctionDistance}, \textit{SpamStorage} and \textit{PartnerSorting} strategy is shown in Fig.~\ref{fig:example_movements} in Appendix \ref{sec:append_example_ion_movements}, and Appendix \ref{appendix:ionorchestrationexample} shows an example that demonstrates the necessity of the \textit{ComputeOrder} heuristic.


\vspace{1 \baselineskip}
This architecture as described above is motivated by our observation that the tasks of Graph Serialization and Ion Orchestration are intrinsically linked and cannot be optimized independently. Using this iterative architecture allows us to tackle this problem, by considering many small, local optimization problems, each dealing with a subset of the circuit (i.e. short graph paths), rather than trying to find the globally optimal serialized circuit.
Furthermore, this design allows for gates to be reordered while at the same time ensuring that the next gate sequence is known during the ion orchestration step, which allows for more efficient move operations. 

A simple example showing how an OpenQASM file is compiled using our backend compilation phases is shown in \cite{tiasm2022compiler}.

\section{Evaluation}
\label{sec:evaluation}

\subsection{Circuit optimization}
In order to see how the template performs, we looked at circuits with 5 qubits and a depth of 40, built up from a random sample of standard Qiskit gates consisting of single qubit, two-qubit and three-qubit gates \cite{Qiskit}. Firstly, \texttt{pytket} optimizations already reduce the number of gates significantly (see also \cite{benchmark}). Then by implementing the circuit identity, we could see another 8-10\% reduction, which contributes to higher overall circuit fidelities.

\begin{figure*}
\floatbox[{\capbeside\thisfloatsetup{capbesideposition={right,top},capbesidewidth=4cm}}]{figure}[\FBwidth]
{\caption{Evaluation of our compiler by counting the number of movements generated when compiling 500 random circuits as explained in the main text. \textit{Left:} Movement count and compile time for varying path length $n_g$ and all heuristics enabled. \textit{Right:} Showing the influence of optimization strategies. ''Random gate'': randomly select next two-qubit gate, ''Random path'': randomly select a circuit path with $n_g=7$, ''Simulate paths'': simulate all paths with $n_g=7$ and select the best one. Next 4 points: successively adding the heuristics explained in Section~\ref{sec:ion_orchestration}.}\label{fig:path_choice_heuristics}}
{\includegraphics[width=0.65\textwidth]{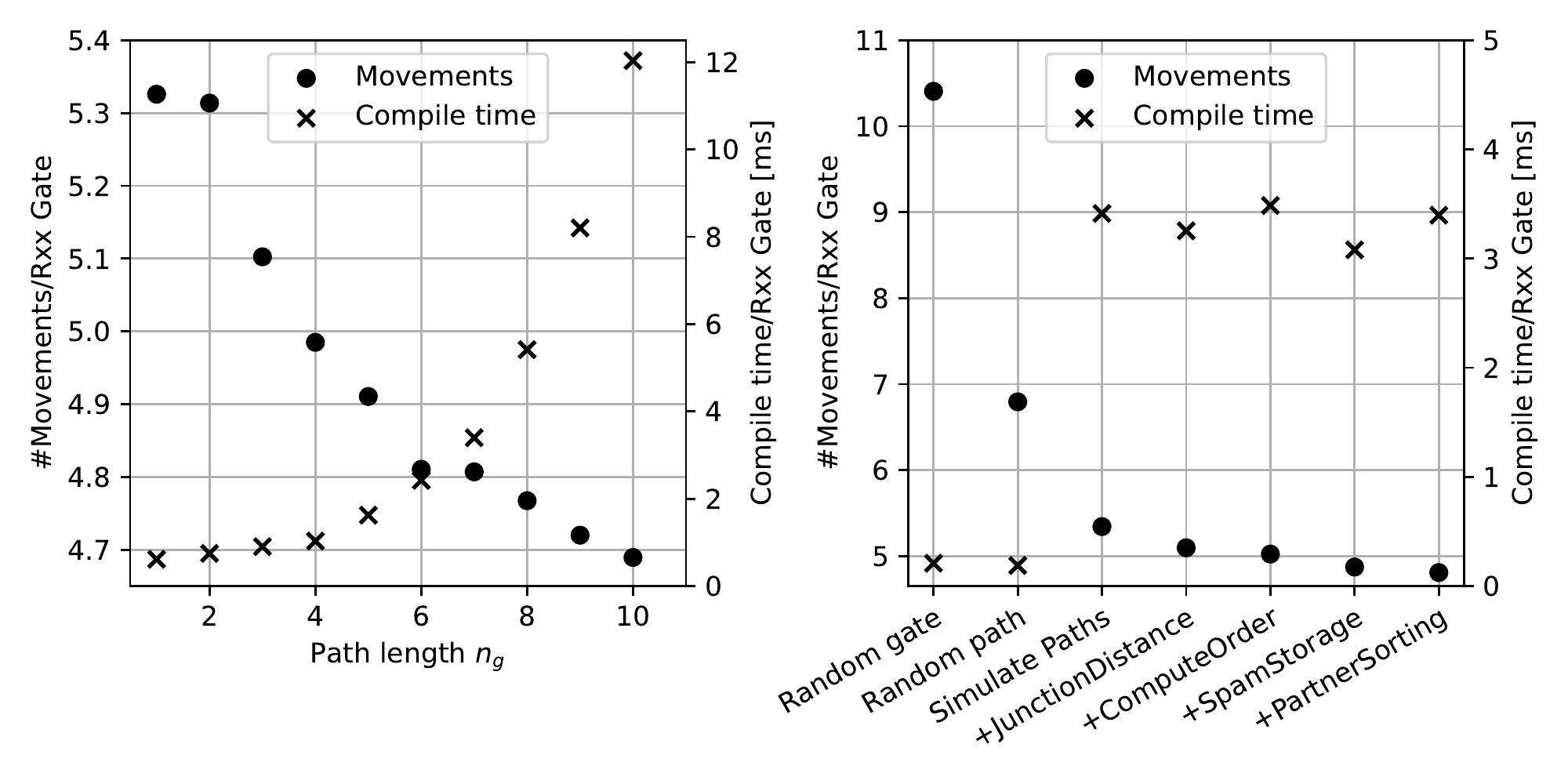}}
\end{figure*}

\subsection{Graph serialization optimizations}
The techniques for serializing the circuit graph are analyzed using a dataset of 500 random circuits on 1-50 qubits, with a mean \tiasm{rxx}-gate count of 52 and a mean \tiasm{rxx}-depth of 20. We compile each circuit and count the total number of movement operations generated. The results are shown in Fig.~\ref{fig:path_choice_heuristics}.

To set these results into context, we try to give an estimate for the number of movements generated by a very naive compiler, that randomly picks the next available two-qubit gate:
The above circuits contain on average 25 qubits. A reasonable equilibrium distribution of ions on the chip for a large random circuit would consist of the \register{STORAGE} and \register{TEMPSTORAGE} areas each containing half of the ions, i.e. around 12 ions per storage register. Therefore, the average ion requires just over 6 movements to be transported to the \register{COMPUTE} area. An execution of a single \tiasm{rxx} gate requires 2 ions to be present in the \register{COMPUTE} area, i.e. around $12$ movements need to be executed per \tiasm{rxx} gate.

If we turn off all heuristics, set $n_g=1$ and randomly choose a circuit path, our compiler best matches the above criteria. With these settings we get $\approx 10.4$ movements/\tiasm{rxx} gate as seen in Fig.~\ref{fig:path_choice_heuristics} (right, ''Random gate''). We do not expect these two values to match exactly as the theoretical expectation is a very crude estimate 
and we have a limited dataset of finite circuits,
however, the above argument gives a rough baseline for the empirical value and can explain the order of magnitude of typical movement counts generated by the compiler.

Next, we point out the fact that merely introducing the circuit path already gives a large reduction in the number of moves. This can be seen by comparing the datapoint ''Random gate'' with ''Random path'' (Fig.~\ref{fig:path_choice_heuristics} (right)) where a random path of length $n_g=7$ is selected. This vast improvement can be explained by the better reuse of ions located in the \register{COMPUTE} area that purely arises from walking along the graph edges instead of picking random gates from the front of the graph.
If one then simulates all graph paths and picks the best one, as for data point ''Simulate paths'' in Fig.~\ref{fig:path_choice_heuristics}, one achieves a further reduction.

The length of the generated paths $n_g$, also has a large influence on the generated number of movements, as evident from Fig.~\ref{fig:path_choice_heuristics} (left). However, increasing $n_g$ has the expected consequence of an exponentially increasing compile time, so for practical applications, $n_g$ should be limited to a fixed, small value.
This highlights the tuning capability of the compiler, allowing it to be adjusted to the required task: for real-time quantum computations the circuit needs to be compiled as fast as possible. However, when a circuit can be compiled beforehand, the quality of the assembly instructions might be of more importance than the compile time.

\subsection{Ion orchestration optimizations}

The heuristics given above for ion orchestration are also analyzed in Fig.~\ref{fig:path_choice_heuristics} (right). These heuristics allow us to  go from 5.34 movements/\tiasm{rxx} gate, to 4.81 movements/\tiasm{rxx} gate for $n_g = 7$, which is significant reduction for circuits with thousands to tens to thousands of \tiasm{rxx} gates.

Overall we are thus able to reduce the $10.4$ moves/\tiasm{rxx} of the naive compiler (Fig.~\ref{fig:path_choice_heuristics} (right, ''Random gate'')) all the way down to less than $4.7$ moves/\tiasm{rxx} (Fig.~\ref{fig:path_choice_heuristics} (left, $n_g = 10$)), which is a reduction by more than 50\,\%.

\section{Related work}
An extensive study of compilers for different QCCD designs was performed in \cite{murali2020}. This work, however, assumes the ability to swap neighbouring ions within the same trap as well as gate applications in all traps. In \cite{sakiMuzzle2021},  optimization techniques based on heuristics were presented for a QCCD with multiple traps arranged in a 1D array. Both works focus on QCCD architectures that are distinctly different from the relevant QVLS-Q1 chip.


The optimization techniques in \cite{sakiMuzzle2021} also first included the usage of a gate dependency graph to optimize the execution order of gates. Building upon this idea we are generating in this work multiple paths through the gate dependency graph to simultaneously optimize the gate execution order and the ion movements.

\section{Conclusion}
In this paper we presented backend compilation phases for trapped-ion QCCD architectures and provided strategies for simultaneously optimizing gate orderings and ion movements during the generation of assembly instructions. 

As for classical compilers the backend was split into a phase optimizing the IR and a phase generating the assembly instructions. For the circuit optimization phase we demonstrated that \texttt{pytket} transpilation methods can be extended to reduce the amount of gates by up to 10\%.

The assembly instructions are collected by simulating the ion movements corresponding to paths through the circuit graph. To obtain a useful tradeoff between compilation time and quality of assembly instructions, we split the tasks of serializing the circuit graph and orchestrating the ions into multiple local optimization problems that we solved using simple heuristics. This novel strategy is applicable to trapped-ion QCCD architectures in general. The heuristics, however, are dependent on the specific hardware. 

To demonstrate all aspects of the backend compiler phases we developed a trapped-ion assembly language (TIASM) for the QVLS-Q1 chip and implemented our backend compiler phases and specific heuristics for this hardware. That way we were able to reduce the amount of move operations by more than 50\,\% compared to randomly executing gates when using our best heuristics for orchestrating the ions. This result highlights how our backend compiler phases together with well suited heuristics for a specific hardware can contribute to reducing the error that accumulates when applying a quantum circuit on a trapped-ion QCCD architecture.

\section{Outlook}
During the graph serialization step we generate circuit paths by starting from the qubits that are accessible with respect to the gate dependency graph. Depending on the current placement of the ions on the chip there might be some obvious paths that are not worth simulating. In future works we will investigate the effect of heuristics during the generation of graph paths 
in order to reduce the number of paths generated, and reducing the exponential complexity of generating the graph paths.

On top of that, we are planning to improve the rating of the graph paths after their simulation. In this work multiple graph paths are simulated and the graph path that resulted in the least amount of movements is chosen. Instead one could investigate different cost functions that not only take into account the amount of movements, but also other criteria like the expected number of future movements based on the placements of the ions after a graph path was applied.

The QVLS-Q1 chip architecture that was used to demonstrate our backend compiler phases can be seen as a fundamental building block of a larger trapped-ion quantum computer. To cope with multiple connected junctions we will extend the compiler in the future to also optimize the ion movements and gate executions in larger architectures. Here the additional challenge will be to decide which gates should be applied sequentially in one building block and which gates could be applied in parallel using the \register{COMPUTE} zones of all building blocks.

\section*{Acknowledgment}
We gratefully thank Ina Schaefer for helpful discussions. This work was funded by the Quantum Valley Lower Saxony (QVLS) through the Volkswagen foundation and the ministry for science and culture of Lower Saxony and by Germany’s Excellence Strategy – EXC-2123 QuantumFrontiers – 390837967.

\printbibliography

\begin{figure*}
    \centering
    \includegraphics[width=0.9\textwidth]{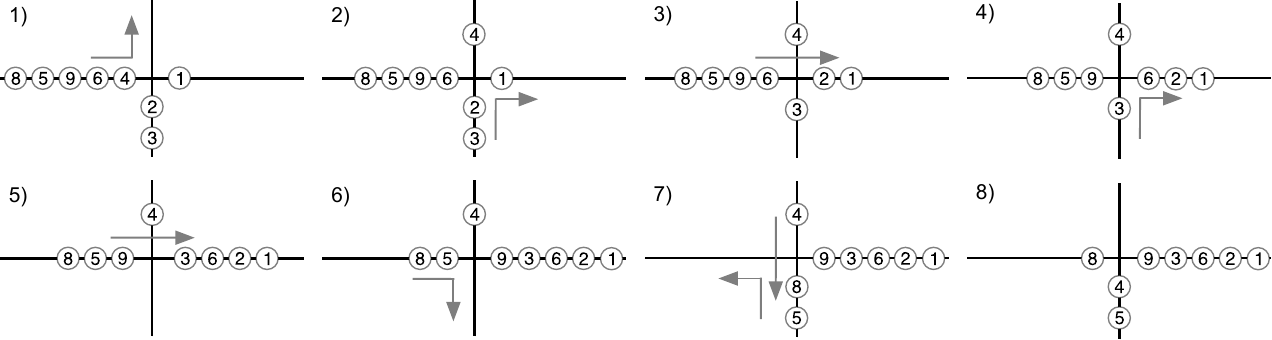}
    \caption{An example of the graph path (\tiasm{RXX}(2, 3), \tiasm{RXX}(5, 8), \tiasm{RXX}(4, 5)) being applied, starting from a possible mid-circuit configuration just after the first \tiasm{RXX} gate of the path has been applied. The arrows indicate the movement operation that needs to be applied to obtain the next diagram. Additionally, ions 3 and 9 are next partners due to an upcoming \tiasm{RXX}(3, 9) gate in the circuit graph. Between steps 1) and 2) the \textit{SpamStorage} heuristic from Section~\ref{sec:ion_orchestration} is made use of, because ion 4 will be needed for an upcoming gate in the graph path. This prevents the ion to be moved unnecessarily far from the junction. Between steps 3) and 5), the \textit{PartnerSorting} heuristic is made use of, as ions 3 and 9 are next partners and they can be placed next to each other, simply by delaying the removal of ion 3 from the \register{COMPUTE} area.
    Note: between 6) and 7), as well as between 7) and 8) two movement operations are applied.}
    \label{fig:example_movements}
\end{figure*}

\appendix
\section{Appendix}

\subsection{Circuit optimization example}
\label{appendix:optimizationexample}
Let us also show an example for the circuit  \hyperref[circ:maslov-templ]{template} that we introduced above. The main advantage of it is that it allows us to replace a particular sequence of the single qubit gates $\tiasm{rx}(\theta_1),\tiasm{ry}(\theta_2),\tiasm{rx}(\theta_1)$, with just one gate $\tiasm{r}(\theta_3, \theta_4)$. For instance, if we find the following sequence of gates in a circuit:
\newline
$$
    \Qcircuit @C=1em @R=.7em {
      & \gate{\tiasm{rx}(\theta_1)} & \gate{\tiasm{ry}(\theta_2)} & \multigate{1}{\tiasm{rxx}(\alpha)} & \gate{\tiasm{rx}(\theta_1)} & \qw \\
      & \qw & \qw & \ghost{\tiasm{rxx}(\alpha)} & \qw & \qw
      }
$$ 
then to be able to use the identity, we need to move the last \tiasm{rx} gate in between the \tiasm{ry} and the \tiasm{rxx} gates. This is possible since \tiasm{rx} and \tiasm{rxx} gates commute with each other. Consequently, the circuit can be reduced to:
$$
    \Qcircuit @C=1em @R=.7em {
      & \gate{\tiasm{r}(\theta_3, \theta_4)} & \multigate{1}{\tiasm{rxx}(\alpha)} & \qw \\
      & \qw & \ghost{\tiasm{rxx}(\alpha)} & \qw
      }
$$ 
Note, that any general \tiasm{r} gate can be directly implemented with the ion-trap quantum hardware at hand, thus the number of gates has been truly reduced. An example where we cannot find the right sequence of gates via commutation is when there is an additional \tiasm{ry} gate in the circuit after the $\tiasm{rxx}$ gate. In this case, the $\tiasm{rx}$ gate cannot commute through the $\tiasm{ry}$ gate, preventing us from using the identity.

\subsection{Graph path generation example}
\label{appendix:graphpathgeneration}

In the following circuit all gates are two-qubit gates ($F$ acts on qubits 2 and 4).

$$
    \Qcircuit @C=1em @R=.7em {
     \lstick{q_0} & \qw & \multigate{1}{\tiasm{c}} &\qw & \qw  & \qw\\
     \lstick{q_1} & \multigate{1}{\tiasm{a}} & \ghost{\tiasm{c}} & \multigate{1}{\tiasm{e}} & \qw  & \qw\\
     \lstick{q_2} & \ghost{\tiasm{a}} & \multigate{1}{\tiasm{d}} & \ghost{\tiasm{e}} & \multigate{2}{\tiasm{f}} & \qw\\
     \lstick{q_3} & \qw & \ghost{\tiasm{d}} & \qw & \ghost{\tiasm{f}} & \qw\\
     \lstick{q_4} & \multigate{1}{\tiasm{b}} & \qw & \qw & \ghost{\tiasm{f}} & \qw\\ 
     \lstick{q_5} & \ghost{\tiasm{b}} & \qw & \qw & \qw & \qw\\  
      }
$$ 

For $n_g = 3$ we generate the following 4 paths through the circuit by following the nodes of the dependency graph:

\begin{enumerate}
    \item[] 1) A, C, D \qquad 2) A, D, C
    \item[] 3) B, A, C \qquad 4) B, A, D
\end{enumerate}

Notice that there are four more paths that satisfy the gate dependencies (e.g. A, B, C). However, we only consider paths that follow the nodes of the dependency graph. Executing gate B after gate A would lead to a complete exchange of the ions in the compute zone, and therefore probably to more move operations.


Following the same strategy we can generate 4 paths for $n_g = 6$:

\begin{enumerate}
    \item[] 1) A, C, D, E, B, F \qquad 2) A, D, C, E, B, F
    \item[]  3) B, A, C, D, E, F \qquad 4) B, A, D, C, E, F
\end{enumerate}

\subsection{Example ion movements}\label{sec:append_example_ion_movements}

Fig.~\ref{fig:example_movements} shows an example of a graph path being applied by moving the ions on the chip, such that they end up in the \register{COMPUTE} area in the correct sequence. For moving the ions, some heuristics from Section~\ref{sec:ion_orchestration} are applied in order to reduce the number of movements.

\subsection{\textit{ComputeOrder} heuristic example}
\label{appendix:ionorchestrationexample}
The following circuit is one of the simplest examples, that highlights a possible optimization regarding ion orchestration.
$$
    \Qcircuit @C=1em @R=.7em {
     \lstick{q_0} & \multigate{1}{\tiasm{rxx}} & \qw & \multigate{1}{\tiasm{rxx}} & \qw \\
     \lstick{q_1} & \ghost{\tiasm{rxx}} & \multigate{1}{\tiasm{rxx}} & \ghost{\tiasm{rxx}} & \qw \\
     \lstick{q_2} & \qw & \ghost{\tiasm{rxx}} & \qw & \qw
      }
$$ 
Assuming ions $q_0$ and $q_1$ are not in the \register{COMPUTE} register and are equally close to the junction (as would for example happen at the beginning of a circuit when no qubits id's have yet been assigned to the ions), a decision has to be made regarding the order in which ions $q_0$ and $q_1$ are put into the \register{COMPUTE} register.
Here it is more efficient, to start with ion 1, as it can remain in \register{COMPUTE} for longer, and does not have to be moved again for the execution of the two subsequent $\tiasm{rxx}$ gates.

\end{document}